\documentclass[a4paper]{report}
\usepackage[utf8]{inputenc}
\usepackage[T1]{fontenc}
\usepackage{RJournal}
\usepackage{amsmath,amssymb,array}
\usepackage{booktabs}

\providecommand{\tightlist}{%
  \setlength{\itemsep}{0pt}\setlength{\parskip}{0pt}}

\usepackage{amsmath}

\begin{document}

\sectionhead{Contributed research article}
\volume{16}
\volnumber{2}
\year{2024}
\month{June}

\begin{article}
\title{\texttt{pencal}: an \texttt{R} Package for the Dynamic Prediction
of Survival with Many Longitudinal Predictors}
\author{by Mirko Signorelli}

\maketitle

\abstract{%
In survival analysis, longitudinal information on the health status of a
patient can be used to dynamically update the predicted probability that
a patient will experience an event of interest. Traditional approaches
to dynamic prediction such as joint models become computationally
unfeasible with more than a handful of longitudinal covariates,
warranting the development of methods that can handle a larger number of
longitudinal covariates. We introduce the \texttt{R} package
\CRANpkg{pencal}, which implements a Penalized Regression Calibration
(PRC) approach that makes it possible to handle many longitudinal
covariates as predictors of survival. \CRANpkg{pencal} uses
mixed-effects models to summarize the trajectories of the longitudinal
covariates up to a prespecified landmark time, and a penalized Cox model
to predict survival based on both baseline covariates and summary
measures of the longitudinal covariates. This article illustrates the
structure of the \texttt{R} package, provides a step by step example
showing how to estimate PRC, compute dynamic predictions of survival and
validate performance, and shows how parallelization can be used to
significantly reduce computing time.
}

\hypertarget{introduction}{%
\section{Introduction}\label{introduction}}

Risk prediction models \citep{steyerberg2009} allow to estimate the
probability that an event of interest will occur in the future. Such
models are commonly employed in the biomedical field to estimate the
probability that an individual will experience an adverse event, and
their output can be used to inform patients, monitor their disease
progression, and guide treatment decisions.

Traditionally, risk prediction models only used covariate values
available at the beginning of the observation period to predict
survival. Thus, predictions based on such models could not exploit
information gathered at later time points. Because information about the
evolution of time-dependent covariates may influence the occurrence of
the survival outcome, it is desirable to be able to dynamically update
predictions of survival as more longitudinal information becomes
available.

Dynamic prediction models employ both baseline and follow-up information
to predict survival, and they can be used to update predictions each
time new follow-up data are gathered. Three commonly-used statistical
methods for dynamic prediction are the time-dependent Cox model, joint
modelling of longitudinal and survival data, and landmarking approaches.

The time-dependent Cox model \citep{therneau2000} is an extension of the
Cox proportional hazards model that allows for the inclusion of
time-dependent covariates. It assumes the value of a time-dependent
covariate to be constant between two observation times, and it is only
suitable for exogenous time-dependent covariates. The model can be
estimated using the \texttt{R} package \CRANpkg{survival}
\citep{therneau2000}.

Joint models for longitudinal and survival outcomes
\citep{henderson2000} are shared random effects models that combine a
submodel for the longitudinal covariates (typically linear mixed models)
and one for the survival outcome (usually a Cox model or a parametric
survival model). Thanks to the shared random effects formulation, such
models are capable to account for a possible interdependence between the
longitudinal covariates and the survival outcome. However, the
estimation of the shared random effects model is a computationally
intensive task that has so far restricted the application of joint
models to problems with one or few longitudinal covariates. Over the
years, several alternative approaches to the estimation of joint models
have been proposed, among which are the \texttt{R} packages \CRANpkg{JM}
\citep{rizopoulos2010}, \CRANpkg{JMbayes} \citep{rizopoulos2014},
\CRANpkg{joineR} \citep{philipson2018} and \CRANpkg{joineRML}
\citep{hickey2018}.

Lastly, landmarking \citep{vanhouwelingen2007} is an approach that
dynamically adjusts predictions by refitting the prediction model using
all subjects that are still at risk at a given landmark time.
Landmarking typically involves two modelling steps. In the first step,
repeated measurements of the time-dependent covariates up to the
landmark time are summarized using either a summary measure or a
suitable statistical model. In the second step, the summaries thus
computed are used as predictors of survival alongside with the
time-independent covariates. The simplest form of landmarking is the
last observation carried forward (LOCF) method, which uses the last
available measurement of each longitudinal covariate taken up to the
landmark time as summary. The main advantage of this approach is that it
is easy to implement, it is computationally straightforward and, thus,
it does not require the development of dedicated software. Limitations
of LOCF landmarking include the fact that it discards all previous
repeated measurements, failing to make efficient use of the available
longitudinal information, and it does not perform any measurement error
correction on the longitudinal covariates (this may be particularly
desirable for biomarkers and diagnostic tests that are typically subject
to measurement error). To overcome these limitations, mixed-effects
models can be used to model the trajectories of the longitudinal
covariates \citep{signorelli2021, putter2022, devaux2023}. While
\citet{putter2022} focused on situations with a single longitudinal
marker, \citet{signorelli2021} and \citet{devaux2023} proposed two
methods, respectively called Penalized Regression Calibration (PRC) and
DynForest, that can deal with a large number of longitudinal covariates.
Notably, the estimation of PRC and DynForest is much more complex than
that of LOCF landmarking, warranting dedicated software that can
facilitate the implementation of such methods. PRC is implemented in the
\texttt{R} package \CRANpkg{pencal} \citep{signorelli2023}, whereas
DynForest in the \texttt{R} package \CRANpkg{DynForest}
\citep{devaux2023}.

In this article we introduce the \texttt{R} package \CRANpkg{pencal},
which implements the PRC approach. PRC uses mixed-effects models to
describe and summarize the longitudinal biomarker trajectories; the
summaries thus obtained are used as predictors of survival in a Cox
model alongside with any relevant time-independent covariate. To account
for the possible availability of a large (potentially high-dimensional)
number of time-independent and longitudinal covariates and to reduce the
risk of overfitting the training data, PRC uses penalized maximum
likelihood to estimate the aforementioned Cox model.

The remainder of the article is organized as follow. In the next section
we describe the dynamic prediction problem, the PRC statistical
methodology \citep{signorelli2021} behind \CRANpkg{pencal}, and the
problem of evaluating the model's predictive performance. Next we
provide a general overview of the \texttt{R} package, discussing the
implementation details of its functions for model estimation, prediction
and performance validation. Furthermore, we provide a step by step
example that shows how to use \CRANpkg{pencal} to implement dynamic
prediction on a real-world dataset that comprises several longitudinal
covariates. We present the results of 4 simulations that assess the
relationship between computing time and sample size, number of
covariates and number of bootstrap samples used to validate model
performance, showing how parallelization may reduce computing time
significantly. Lastly, we provide some final remarks, and discuss
limitations and possible extensions of the current approach.

\hypertarget{statistical-methods}{%
\section{Statistical methods}\label{statistical-methods}}

\hypertarget{input-data-and-notation}{%
\subsection{Input data and notation}\label{input-data-and-notation}}

We consider a setting where \(n\) subjects are followed from time
\(t = 0\) until an event of interest occurs. For each subject
\(i \in \{1, ..., n\}\) we observe the pair \((t_i, \delta_i)\), where
\(\delta_i\) is a dummy variable that indicates whether the event is
observed at time \(t = t_i\) (\(\delta_i = 1\)), or the observation of
the event is right-censored at \(t = t_i\) (\(\delta_i = 0\)). Thus,
\(t_i\) corresponds to the survival time if \(\delta_i = 1\), and to the
censoring time if \(\delta_i = 0\).

In addition to \((t_i, \delta_i)\), we assume that both baseline and
follow-up information is collected from the same subjects, and that the
number of variables gathered may be large. We consider a flexible
unbalanced study design where the number and timing of the follow-up
times can differ across subjects. We denote by \(m_i \geq 1\) the number
of repeated measurements available for subject \(i\), and by
\(t_{i1}, ..., t_{i m_i} \:\: (t_{ij} \geq 0 \:\: \forall j)\) the
corresponding follow-up times. For each subject \(i\) we observe:

\begin{enumerate}
\def\labelenumi{\arabic{enumi}.}
\tightlist
\item
  a vector of \(k\) baseline (time-fixed) predictors
  \(x_i = \left( x_{1i}, ..., x_{ki} \right)\);
\item
  \(m_i\) vectors of \(p\) longitudinal (time-varying) predictors
  \(y_{ij} = \left( y_{1ij}, ..., y_{pij} \right), \:\: j = 1,..., m_i\)
  measured at times \(t_{i1}, ..., t_{im_i}\). Note that not all
  longitudinal predictors ought to be measured at every follow-up time
  \(t_{ij}\), and the number of available measurements is thus allowed
  to differ across longitudinal predictors.
\end{enumerate}

\hypertarget{dynamic-prediction-of-survival}{%
\subsection{Dynamic prediction of
survival}\label{dynamic-prediction-of-survival}}

Let \(S_i(t) = P(T_i > t)\) denote the survival function, i.e., the
probability that subject \(i\) has not experienced the event up to time
\(t\), and let
\(S_i(t_B | t_A) = P(T_i > t_B | T_i > t_A), \: t_B \geq t_A\) denote
the conditional probability that subject \(i\) survives up until
\(t_B\), given that they survived up until \(t_A\). Our goal is to
predict the probability of survival of subject \(i\) given all the
available information up until a given landmark time \(t_L > 0\),
namely:

\begin{equation}
S_i(t | t_L, x_i, \mathcal{Y}_i(t_L)) =  P(T_i > t | T_i > t_L, x_i, \mathcal{Y}_i(t_L)),
\label{eq:cond_surv}
\end{equation}

\noindent where
\(\mathcal{Y}_i(t_L) = \{ y_{i1}, ..., y_{ir}: t_{ir} \leq t_L \}\)
denotes all repeated measurements available up to the landmark time for
subject \(i\).

In practice, often one may be interested in computing the predictions of
survival in \eqref{eq:cond_surv} over a range of \(q\) landmark times
\(t_{L1}, t_{L2}, ..., t_{Lq}\). By doing so, information from more and
more repeated measurements can be incorporated in the prediction model
as time passes, and predictions can be updated based on the latest
available information. This approach is referred to as \emph{dynamic
prediction}, as it involves dynamic updates of model estimates and
predictions over time. In this article we illustrate how to use
\CRANpkg{pencal} to compute predictions of the conditional survival
probabilities in \eqref{eq:cond_surv} for a single landmark time
\(t_L\). Note that implementing a dynamic prediction approach with
\CRANpkg{pencal} is straightforward, as for each landmark time
\(t_{Ls}\) one simply needs to refit PRC over a dataset that comprises
all repeated measurements available up until \(t = t_{Ls}\) for the
subjects that survived up until the landmark time \(t_{Ls}\) (i.e.,
including subject \(i\) if and only if \(t_i \geq t_{Ls}\)).

\hypertarget{penalized-regression-calibration}{%
\subsection{Penalized regression
calibration}\label{penalized-regression-calibration}}

PRC \citep{signorelli2021} is a statistical method that makes it
possible to estimate the conditional survival probabilities in
\eqref{eq:cond_surv} using \(x_i\) and \(\mathcal{Y}_i(t_L)\) as inputs.
The estimation of PRC requires a multi-step procedure that comprises the
3 following steps:

\begin{enumerate}
\def\labelenumi{\arabic{enumi}.}
\tightlist
\item
  model the evolution over time of the longitudinal predictors in
  \(\mathcal{Y}_i(t_L)\) using linear mixed models (LMM,
  \citet{mcculloch2004}) or multivariate latent process mixed models
  (MLPMM, \citet{proust2013});
\item
  use the model(s) fitted at step 1 to compute summaries of the
  trajectories described by the longitudinal predictors (i.e., the
  predicted random effects);
\item
  estimate a penalized Cox model for the survival outcome
  \((t_i, \delta_i)\) using as covariates both the baseline predictors
  \(x_i\) and the predicted random effects computed in step (2).
\end{enumerate}

For simplicity, in this section we describe the version of PRC where in
step 1 each longitudinal covariate is modelled using a separate LMM.
This version of the model is referred to as PRC LMM in
\citet{signorelli2021}. Note that alongside with the PRC LMM approach,
\citet{signorelli2021} also proposed a second approach called PRC MLPMM,
where groups of longitudinal covariates are modelled jointly using the
MLPMM. This alternative approach can be of interest when multiple
longitudinal items are employed to measure the same underlying quantity
(for example, multiple antibodies that target the same protein). For the
formulation of the PRC MLPMM approach, we refer readers to Sections
2.1-2.3 of \citet{signorelli2021} as the notation for steps 1 and 2
using the MLPMM is significantly more involved.

Denote by \(\mathcal{I}(t_L) = \{ i: t_i > t_L \}\) the set of subjects
that survived up until the landmark time \(t_L\). Let
\(y_{si} = (y_{si1}, ..., y_{sir})\), where \(t_{ir} \leq t_L\) denotes
the last follow-up time before \(t_L\) for subject \(i\), be the vector
that comprises all the measurements of the \(s\)-th longitudinal
variable \(Y_s\) available up to the landmark time. In the first step of
PRC, we model the evolution over time of each longitudinal covariate
\(Y_s\) through a linear regression model

\begin{equation}
y_{si} = W_{si} \beta_s + Z_{si} u_{si} + \varepsilon_{si}, \: i \in \mathcal{I}(t_L),
\label{eq:lmm}
\end{equation}

where \(\beta_s\) is a vector of fixed effect parameters,
\(u_{si} \sim N(0, D_s)\) is a vector of random effects,
\(\varepsilon_{si} \sim N(0, \sigma^2_s I_{m_i})\) is the error term
vector, and \(W_{si}\) and \(Z_{si}\) are design matrices associated to
\(\beta_s\) and \(u_{si}\). As an example, later in this article we will
consider an example where we let \(y_{si}\) depend on the age \(a_{ij}\)
of subject \(i\) at each visit, and include a random intercept and
random slope in the LMM:

\begin{equation}
y_{sij} = \beta_{s0} + u_{si0} + \beta_{s1} a_{ij} + u_{si1} a_{ij} + \varepsilon_{sij},
\label{eq:lmmwithslope}
\end{equation}

where \((u_{s0}, u_{s1})^T \sim N(0, D_s)\) is a vector of random
effects that follows a bivariate normal distribution. We employ maximum
likelihood (ML) estimation to estimate \(\beta_s\), \(D_s\) and
\(\sigma^2_s\) in model \eqref{eq:lmm}.

In the second step of PRC, we use the ML estimates from step 1 to derive
summaries of the individual longitudinal trajectories for each
biomarker. These are the predicted random effects, which can be computed
as

\begin{equation}
\hat{u}_{si} =E(u_{si} | Y_{si} = y_{si}) = 
\hat{D}_s Z_{si}^T \hat{V}_{si}^{-1} (y_{si} - X_{si} \hat{\beta}_s),
\label{eq:predranef}
\end{equation}

where \(\hat{V}_{si} = Z_{si} \hat{D}_s Z_{si}^T + \hat{\sigma}^2_s I\).

In the third step of PRC, we model the relationship between the survival
outcome and the baseline and longitudinal predictors. This is achieved
through the specification of a Cox model where we include the baseline
predictors \(x_i\) and the summaries of the longitudinal predictors
\(\hat{u}_i = (\hat{u}_{1i}, ..., \hat{u}_{pi})\) as covariates:

\begin{equation}
h(t_i | x_i, \hat{u}_i) = h_0(t_i) \exp \left( \gamma x_i + \delta \hat{u}_i \right),
\label{eq:cox}
\end{equation}

where \(h_0(t_i)\) is the baseline hazard function, and \(\gamma\) and
\(\delta\) are vectors of regression coefficients. Since our approach
allows for the inclusion of a potentially large number of baseline and
longitudinal covariates, we estimate model \eqref{eq:cox} using
penalized maximum likelihood (PML, \citet{verweij1994}). As penalties we
consider the ridge (\(L_2\)), lasso (\(L_1\)) and elasticnet penalties.
The elasticnet penalty for model \eqref{eq:cox} is given by

\begin{equation}
\lambda  \left[
\alpha \left( \sum_{s=1}^p |\gamma_s| + \sum_{s=1}^p |\delta_s| \right)
+ (1 - \alpha) \left( \sum_{s=1}^p \gamma_s^2 + \sum_{s=1}^p \delta_s^2 \right)
\right],
\label{eq:elasticnet}
\end{equation}

where \(\lambda \geq 0\) and \(\alpha \in [0, 1]\). The ridge penalty is
obtained by setting \(\alpha = 0\), and the lasso penalty by fixing
\(\alpha = 1\).

\hypertarget{computation-of-the-predicted-survival-probabilities}{%
\subsection{Computation of the predicted survival
probabilities}\label{computation-of-the-predicted-survival-probabilities}}

Once models \eqref{eq:lmm} and \eqref{eq:cox} have been estimated, the
predicted survival probabilities
\(S_i(t | t_L, x_i, \mathcal{Y}_i(t_L))\) are computed using

\begin{equation}
\hat{S}_i(t | t_L, x_i, \mathcal{Y}_i(t_L)) =
\exp \left( 
- \int_0^t \hat{h}_0(s) \exp( \hat{\gamma} x_i + \hat{\delta} \hat{u}_i )
\right),
\label{eq:predsurv}
\end{equation}

where \(\hat{h}_0(s)\) is the estimated baseline hazard function,
\(\hat{\gamma}\) and \(\hat{\delta}\) are the PML estimates of
\(\gamma\) and \(\delta\) obtained in step 3, and \(\hat{u}_i\) contains
the predicted random effects computed in step 2.

For subjects \(i \in \mathcal{I}(t_L)\), who are included in the
training set and survived up until \(t_L\), computation of
\eqref{eq:predsurv} is straightforward, since the predicted random
effects \(\hat{u}_i\) for such subjects have already been computed in
step 2. Predictions of \(S_i(t | t_L, x_i, \mathcal{Y}_i(t_L))\) for a
new subject \(i = n + 1\) who survived up until \(t_L\), but was not
part of the training set is a bit more complex: before computing
\eqref{eq:predsurv}, one first needs to compute the predicted random
effects for this new subject using \eqref{eq:predranef}. Note that such
computation is feasible if and only if measurements of both baseline and
longitudinal covariates (up to \(t_L\)) are available for this new
subject.

\hypertarget{evaluation-of-the-predictive-performance}{%
\subsection{Evaluation of the predictive
performance}\label{evaluation-of-the-predictive-performance}}

We consider the time-dependent area under the ROC curve (tdAUC,
\citet{heagerty2000}), the concordance index or C index
\citep{pencina2004} and the Brier score \citep{graf1999} as measures of
predictive performance. To obtain unbiased estimates of these
performance measures, \citet{signorelli2021} proposed a cluster
bootstrap optimism correction procedure (CBOCP) that generalizes the use
of the bootstrap as internal validation method to problems involving
repeated measurement data. As an alternative to the CBOCP, one may
choose to implement a cross-validation approach instead. Should the user
opt for such an alternative, we recommend the use of repeated
cross-validation over simple cross-validation to achieve a level of
accuracy comparable to that of the CBOCP.

\hypertarget{the-r-package-pencal}{%
\section{\texorpdfstring{The \texttt{R} package
\texttt{pencal}}{The R package pencal}}\label{the-r-package-pencal}}

In this Section we introduce the functions for the estimation of PRC,
the computation of the predicted survival probabilities and the
validation of predictive performance, providing an overview of the
relevant estimation approaches and some important implementation
details.

Table \ref{tab:functions} provides a side-by-side overview of the
functions that can be used to implement the PRC LMM and PRC MLPMM
approaches. Note that while two different functions (one for each
approach) are needed for the three estimation steps and the computation
of the survival probabilities, the evaluation of the predictive
performance is implemented in a single function that works with inputs
from both approaches.

\begin{table*}
\caption{Overview of the \texttt{pencal} functions that implement the different modelling steps for the PRC LMM and PRC MLPMM approaches. \label{tab:functions}}
\centering
\begin{tabular}{l|c|c}
Task & PRC LMM & PRC MLPMM\\
\hline
Step 1: estimate the mixed-effects models & \texttt{fit\_lmms} & \texttt{fit\_mlpmms} \\ 
Step 2: compute the predicted random effects & \texttt{summarize\_lmms} & \texttt{summarize\_mlpmms} \\ 
Step 3: estimate the penalized Cox model & \texttt{fit\_prclmm} & \texttt{fit\_prcmlpmm} \\
Computation of predicted survival probabilities & \texttt{survpred\_prclmm} & \texttt{survpred\_prcmlpmm} \\
Evaluation of predictive performance & \texttt{performance\_prc} & \texttt{performance\_prc}\\
\end{tabular}
\end{table*}

\hypertarget{model-estimation-and-prediction}{%
\subsection{Model estimation and
prediction}\label{model-estimation-and-prediction}}

The first step of PRC involves the estimation of mixed-effects models
for the longitudinal outcomes. For the PRC LMM approach, this can be
done through the \texttt{fit\_lmms} function, that proceeds to estimate
\(p\) LMMs (one LMM for each of the longitudinal outcomes). Estimation
of the LMMs is performed by maximum likelihood through the \texttt{lme}
function from the \texttt{R} package \CRANpkg{nlme}
\citep{pinheiro2000}. For the PRC MLPMM approach, the first step
involves estimating one MLPMM for each group of longitudinal covariates.
Estimation of the MLPMMs is done by maximum likelihood using the
modified Marquardt algorithm described in \citet{proust2013}, as
implemented in the \texttt{multlcmm} from the \texttt{R} package
\CRANpkg{lcmm} \citep{proust2017}.

The second step of PRC requires the computation of the predicted random
effects. The function \texttt{summarize\_lmms} implements this for the
PRC LMM approach. The function takes the output of \texttt{fit\_lmms} as
input, and proceeds to the computation of the predicted random effects
using equation \eqref{eq:predranef}. Similarly, the function
\texttt{summarize\_mlpmms} does the same for the PRC MLPMM approach by
taking the output of \texttt{fit\_mlpmms} as input and computing the
predicted random effects using the formula given in equation (4) of
\citet{signorelli2021}.

The third step of PRC requires the estimation of a Cox model where the
baseline covariates and the predicted random effects are used as
covariates.Estimation of such model can be performed using the function
\texttt{fit\_prclmm} for the PRC LMM approach and \texttt{fit\_prcmlpmm}
for the PRC MLPMM approach. These functions proceed to the estimation of
the aforementioned Cox model by penalized maximum likelihood through the
function \texttt{cv.glmnet} from the \texttt{R} package \CRANpkg{glmnet}
\citep{simon2011}. If the user chooses the ridge or lasso penalty, then
the selection of the value of the tuning parameter \(\lambda\) is
performed through cross-validation as implemented in \CRANpkg{glmnet}.
If, instead, the elasticnet penalty is used, \texttt{fit\_prclmm} and
\texttt{fit\_prcmlpmm} proceed to perform a nested cross-validation
procedure to jointly select the optimal values of the tuning parameters
\(\alpha\) and \(\lambda\).

Lastly, the function \texttt{survpred\_prclmm} can be used to compute
the predicted survival probabilities as described in equation
\eqref{eq:predsurv} for the PRC LMM approach. The corresponding function
for the PRC MLPMM approach is \texttt{survpred\_prcmlpmm}.

\hypertarget{computation-of-the-cbocp}{%
\subsection{Computation of the CBOCP}\label{computation-of-the-cbocp}}

In \texttt{pencal}, the evaluation of the predictive performance of the
fitted model is done by estimating the tdAUC, C index and Brier score
through the CBOCP described in \citet{signorelli2021}. For both the PRC
LMM and PRC MLPMM approaches, this can be done through the function
\texttt{performance\_prc}. The estimates of the tDAUC, C index and Brier
score are computed using functions from the packages
\CRANpkg{survivalROC} \citep{heagerty2022}, \BIOpkg{survcomp}
\citep{schroder2011} and \CRANpkg{riskRegression} \citep{gerds2022},
respectively.

The computation of the CBOCP requires the choice of the number of
bootstrap replicates \(B\) over which the model should be refitted. This
can be done by specifying the argument \texttt{n.boots} inside the
functions that implement the first step of PRC, namely
\texttt{fit\_lmms} for the PRC LMM approach and \texttt{fit\_mlpmms} for
the PRC MLPMM one. The supplied value of \texttt{n.boots} is stored in
the output of such functions, and all subsequent functions inherit this
value, automatically performing the computations necessary for the
CBOCP.

If \texttt{n.boots\ =\ 0} (default), the CBOCP is not computed, and
\texttt{performance\_prc} only returns the naïve estimates of predictive
performance. Values of \texttt{n.boots} \(\geq 1\) will trigger the
computation of the CBOCP, and the output of \texttt{performance\_prc}
will additionally include the estimates of the optimism and the
optimism-corrected performance measures. A typical value for \(B\) is
100, but in general we recommend setting \(B\) to a value between 50 and
200 (depending on computing time and the desired level of accuracy, one
may also consider larger values of \(B\)).

\hypertarget{user-friendly-parallelization}{%
\subsection{User-friendly
parallelization}\label{user-friendly-parallelization}}

The computation of the CBOCP is by nature repetitive, as it requires to
repeat steps 1, 2 and 3 over \(B\) bootstrap samples, and to compute
predictions and performance measures both on the bootstrap sample and on
the original dataset to estimate the optimism. Due to this
repetitiveness, such computations can be easily parallelized to reduce
computing time. Moreover, also the estimation of the \(p\) LMMs / MLPMMs
in step 1 can be trivially parallelized.

With the goal of making it as easy as possible for users to parallelize
such computations, the functions in Table \ref{tab:functions}
automatically parallelize the aforementioned computations using the
\texttt{\%dopar\%} operator from the \texttt{R} package
\CRANpkg{foreach} \citep{foreach2022}. The user only needs to specify
the number of cores they want to use for the computation using the
argument \texttt{n.cores}; \CRANpkg{pencal} will automatically take care
of the parallelization.

\hypertarget{classes-methods-and-further-functionalities}{%
\subsection{Classes, methods and further
functionalities}\label{classes-methods-and-further-functionalities}}

Besides the core functions introduced in Table \ref{tab:functions},
\CRANpkg{pencal} comprises additional functions that are shortly
described hereafter.

Three functions are used to simulate the data that are used in the
package documentation to illustrate typical usage of \CRANpkg{pencal}'s
functions. The function \texttt{simulate\_t\_weibull} is used to
generate survival times from a Weibull distribution using the inverse
transformation method. The functions \texttt{simulate\_prclmm\_data} and
\texttt{simulate\_prcmlpmm\_data} are used to generate data for the
estimation of PRC LMM and PRC MLPMM, respectively.

S3 classes and methods are implemented for the outputs of each modelling
step:

\begin{itemize}
\tightlist
\item
  step 1: \texttt{fit\_lmms} and \texttt{fit\_mlpmms} return objects of
  class \texttt{lmmfit} and \texttt{mplmmfit}, respectively. As the
  number of longitudinal covariates increases, step 1 of PRC will
  involve the estimation of many mixed-effects models: to simplify the
  extraction of the estimates of each mixed model, we provide two
  summary functions, \texttt{summary.lmmfit} and
  \texttt{summary.mlpmmfit};
\item
  step 2: \texttt{summarize\_lmms} and \texttt{summarize\_mlpmms} return
  objects of class \texttt{ranefs}, for which a \texttt{summary.ranefs}
  function is available;
\item
  step 3: \texttt{fit\_prclmm} outputs an object of class
  \texttt{prclmm}, and \texttt{fit\_prcmlpmm} one of class
  \texttt{prcmlpmm}. For both classes, \texttt{summary} methods
  (\texttt{summary.prclmm} and \texttt{summary.mlpmmfit}, respectively)
  are implemented.
\end{itemize}

Lastly, it may be sometimes of interest to compare the performance of
PRC to that of either a penalized Cox model that only uses baseline
values of all covariates, or a penalized Cox model with LOFC
landmarking. The \texttt{pencox\_baseline} function provides an
interface to estimate these two models and to compute the associated
CBOCP. Its output can be fed to the function
\texttt{performance\_pencox\_baseline} to obtain the naïve and
optimism-corrected estimates of the tdAUC, C index and Brier score for
these two models.

\hypertarget{dynamic-prediction-with-pencal-a-step-by-step-example}{%
\section{\texorpdfstring{Dynamic prediction with \texttt{pencal}: a step
by step
example}{Dynamic prediction with pencal: a step by step example}}\label{dynamic-prediction-with-pencal-a-step-by-step-example}}

\hypertarget{loading-pbc2data}{%
\subsection{\texorpdfstring{Loading
\texttt{pbc2data}}{Loading pbc2data}}\label{loading-pbc2data}}

To illustrate how to use \CRANpkg{pencal} in practice, we employ data
from a study from a clinical trial on primary biliary cholangitis (PBC)
conducted by the Mayo Clinic from 1974 to 1984 \citep{murtaugh1994}. The
trial recorded the first of two survival outcomes, namely liver
transplantation or death. In this example we focus our attention on the
prediction of deaths, treating patients who underwent liver
transplantation as right-censored. The data are available in
\CRANpkg{pencal} in a list called \texttt{pbc2data} that can be loaded
as follows:

\begin{Schunk}
\begin{Sinput}
library(pencal)
data(pbc2data)
ls(pbc2data)
\end{Sinput}
\begin{Soutput}
#> [1] "baselineInfo"     "longitudinalInfo"
\end{Soutput}
\end{Schunk}

\texttt{pbc2data} contains two data frames: \texttt{baselineInfo}
records the survival information \((t_i, \delta_i)\) and baseline
covariates \(x_i\), whereas \texttt{longitudinalInfo} contains repeated
measurements of the longitudinal predictors \(y_i\). For simplicity, we
rename the two data frames as \texttt{sdata} and \texttt{ldata}:

\begin{Schunk}
\begin{Sinput}
sdata = pbc2data$baselineInfo
ldata = pbc2data$longitudinalInfo
\end{Sinput}
\end{Schunk}

\hypertarget{input-data-format}{%
\subsection{Input data format}\label{input-data-format}}

As detailed in the Statistical Methods section, estimation of PRC
requires the following input data for each subject \(i = 1, ..., n\):

\begin{itemize}
\tightlist
\item
  a pair \((t_i, \delta_i)\) providing the survival outcome for subject
  \(i = 1\);
\item
  a vector of \(k\) baseline covariates \(x_i\);
\item
  an \(m_i \times p\) matrix containing all repeated measurements of the
  \(p\) longitudinal predictors. Within this matrix each column
  corresponds to a predictor, and each row to the measurements
  \(y_{ij}\) collected at time \(t_{ij}, \: j \in \{1, ..., m_i \}\) for
  subject \(i\);
\item
  any longitudinal covariate needed to construct the design matrices
  \(W_{si}\) and \(Z_{si}\) that will be used to estimate the LMMs of
  equation \eqref{eq:lmm}.
\end{itemize}

Such data should be provided to \CRANpkg{pencal} using two data frames.
The first data frame is a data frame that contains information on the
survival outcomes \((t_i, \delta_i)\) and the baseline covariates
\(x_i\). For the PBC2 example, this is the \texttt{sdata} data frame
created above:

\begin{Schunk}
\begin{Sinput}
head(sdata)
\end{Sinput}
\begin{Soutput}
#>    id      time event baselineAge    sex treatment
#> 1   1  1.095170     1    58.76684 female D-penicil
#> 3   2 14.152338     0    56.44782 female D-penicil
#> 12  3  2.770781     1    70.07447   male D-penicil
#> 16  4  5.270507     1    54.74209 female D-penicil
#> 23  5  4.120578     0    38.10645 female   placebo
#> 29  6  6.853028     1    66.26054 female   placebo
\end{Soutput}
\end{Schunk}

This data frame should comprise at least 3 variables: a variable named
\texttt{id} that contains subject identifiers, a variable named
\texttt{time} containing the values of \(t_i\), and a dummy variable
named \texttt{event} that corresponds to \(\delta_i\)
(\texttt{event\ =\ 1} for subjects who experience the event at \(t_i\),
and event = 0 for right-censored observations). Additionally, it can
also contain the baseline covariates \(x_i\) (if any); in the example we
have \(k = 3\) baseline covariates: \texttt{baselineAge}, \texttt{sex}
and \texttt{treatment}.

The second data frame is a dataset in long format that contains the
repeated measurements of the longitudinal predictors \(y_{ij}\) and of
any covariate needed to create the design matrices \(W_{si}\) and
\(Z_{si}\). In our example application, such information is stored in
\texttt{ldata}:

\begin{Schunk}
\begin{Sinput}
head(ldata)
\end{Sinput}
\begin{Soutput}
#>   id      age   fuptime serBilir serChol albumin alkaline  SGOT platelets
#> 1  1 58.76684 0.0000000     14.5     261    2.60     1718 138.0       190
#> 2  1 59.29252 0.5256817     21.3      NA    2.94     1612   6.2       183
#> 3  2 56.44782 0.0000000      1.1     302    4.14     7395 113.5       221
#> 4  2 56.94612 0.4983025      0.8      NA    3.60     2107 139.5       188
#> 5  2 57.44716 0.9993429      1.0      NA    3.55     1711 144.2       161
#> 6  2 58.55054 2.1027270      1.9      NA    3.92     1365 144.2       122
#>   prothrombin
#> 1        12.2
#> 2        11.2
#> 3        10.6
#> 4        11.0
#> 5        11.6
#> 6        10.6
\end{Soutput}
\end{Schunk}

This ``longitudinal'' data frame should contain the following
information:

\begin{itemize}
\item
  a variable named \texttt{id} that contains subject identifiers;
\item
  a variable containing the time from baseline \(t_{ij}\) at which the
  measurement was collected. In \texttt{ldata}, we called this variable
  \texttt{fuptime} (short for: follow-up time). Notice that if the
  longitudinal covariates are measured at \(t_{ij} = 0\), a row with
  \texttt{fuptime\ =\ 0} must be included in \texttt{ldata};
\item
  the \(p\) longitudinal predictors (\texttt{serBilir},
  \texttt{serChol}, \texttt{albumin}, \texttt{alkaline}, \texttt{SGOT},
  \texttt{platelets} and \texttt{prothrombin} in the example);
\item
  the covariates needed to construct \(W_{si}\) and \(Z_{si}\)
  (\texttt{age} in the example).
\end{itemize}

\hypertarget{choice-of-the-landmark-time-and-data-preparation}{%
\subsection{Choice of the landmark time and data
preparation}\label{choice-of-the-landmark-time-and-data-preparation}}

Before proceeding with the estimation of PRC and the computation of the
predicted survival probabilities
\(S_i(t | t_L, x_i, \mathcal{Y}_i(t_L))\), three preliminary steps are
needed. The first involves the choice of the landmark time \(t_L\).
Hereafter we choose \(t_L = 2\) years, meaning that we want to predict
the survival probability \(S(t | t_L = 2, x_i, \mathcal{Y}_i(2))\) for a
subject with baseline covariates \(x_i\) and longitudinal covariates
measured up until two years from baseline \(\mathcal{Y}_i(2)\).

\begin{Schunk}
\begin{Sinput}
# set the landmark time
lmark = 2
\end{Sinput}
\end{Schunk}

Once \(t_L\) has been chosen, the second step is to retain for analysis
only those subjects that survived up until the landmark time, i.e.~all
\(i: t_i \geq t_L\):

\begin{Schunk}
\begin{Sinput}
# remove subjects who had event / were censored before landmark
sdata = subset(sdata, time > lmark)
ldata = subset(ldata, id %in% sdata$id)
\end{Sinput}
\end{Schunk}

Lastly, only repeated measurements taken up to the landmark time
(\(t_{ij} \leq t_L\)) should be retained for modelling, whereas
measurements taken after the landmark time (\(t_{ij} > t_L\)) should be
discarded:

\begin{Schunk}
\begin{Sinput}
# remove measurements taken after landmark:
ldata = subset(ldata, fuptime <= lmark)
\end{Sinput}
\end{Schunk}

\hypertarget{descriptive-statistics-data-visualization-and-transformation}{%
\subsection{Descriptive statistics, data visualization and
transformation}\label{descriptive-statistics-data-visualization-and-transformation}}

After choosing \(t_L = 2\) as landmark time, the number of subjects
retained for model estimation is 278, of which 107 experience the event
of interest, whereas the remaining 171 are right-censored:

\begin{Schunk}
\begin{Sinput}
# number of subjects retained in the analysis:
nrow(sdata)
\end{Sinput}
\begin{Soutput}
#> [1] 278
\end{Soutput}
\begin{Sinput}
# number of events (1s) and censored observations (0s):
table(sdata$event)
\end{Sinput}
\begin{Soutput}
#> 
#>   0   1 
#> 171 107
\end{Soutput}
\end{Schunk}

The estimated survival probability can be visualized through the
Kaplan-Meier estimator in Figure \ref{fig:kaplan} as shown below:

\begin{Schunk}
\begin{Sinput}
library(survival)
library(survminer)
surv.obj = Surv(time = sdata$time, event = sdata$event)
KM = survfit(surv.obj ~ 1, type = "kaplan-meier")
ggsurvplot(KM, data = sdata, risk.table = TRUE, cumevents = TRUE, legend = 'none')
\end{Sinput}
\begin{figure}[htbp]

{\centering \includegraphics[width=5in]{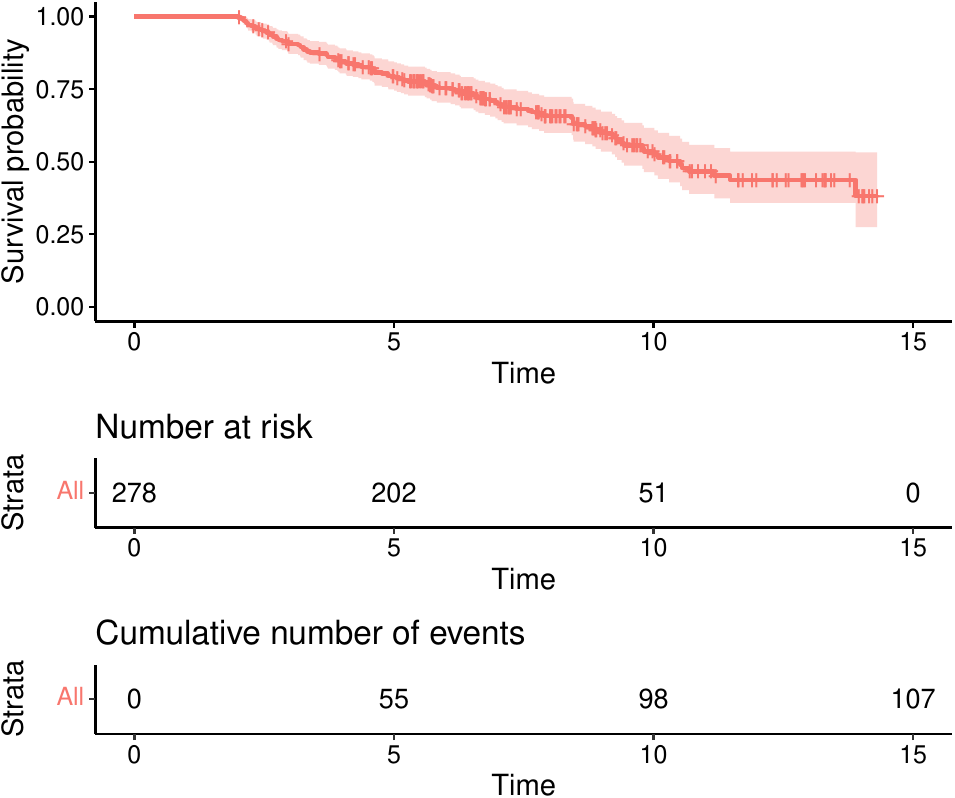} 

}

\caption[Chart displaying the Kaplan-Meier estimates of the conditional survival probability $S(t | 2)$]{Chart displaying the Kaplan-Meier estimates of the conditional survival probability $S(t | 2)$.}\label{fig:kaplan}
\end{figure}
\end{Schunk}

Spaghetti plots displaying the trajectory described by a longitudinal
covariate, as well as density plots to visualize its marginal
distribution, can be created in \texttt{ggplot2} style using:

\begin{Schunk}
\begin{Sinput}
library(ggplot2)
library(gridExtra)
traj1 = ggplot(ldata, aes(x = age, y = serBilir, group = id)) + 
  geom_line(color = 'darkgreen') + theme_classic() + ggtitle('Trajectories of serBilir')
dens1 = ggplot(ldata, aes(x = serBilir)) + 
  geom_density(adjust=1.5, alpha=.4, fill = 'orange') + theme_classic() + 
  ggtitle('Density of serBilir')
grid.arrange(traj1, dens1, ncol = 2)
\end{Sinput}
\begin{figure}[htbp]

{\centering \includegraphics[width=5in]{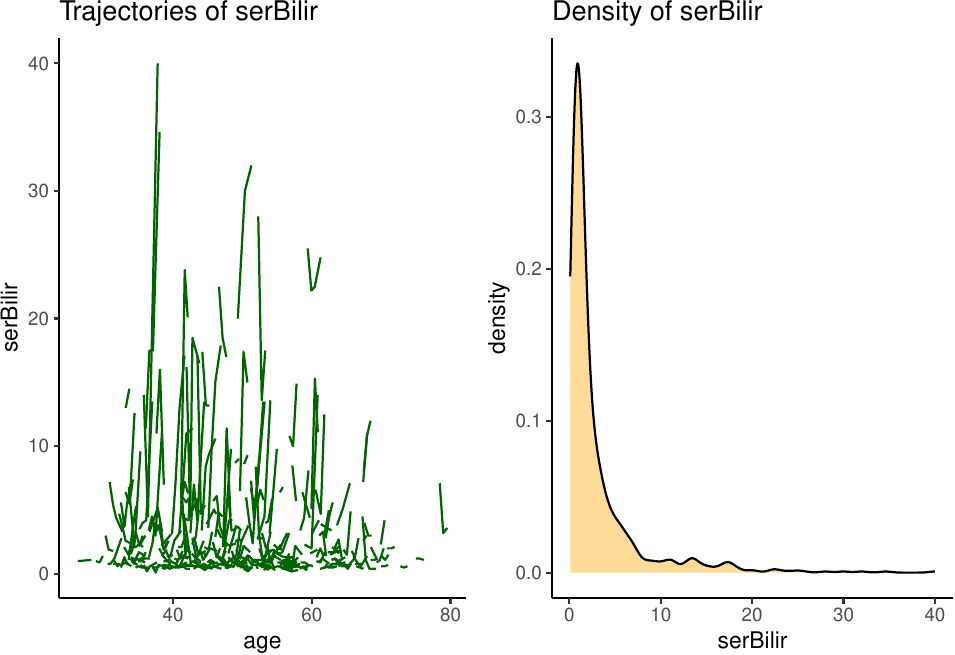} 

}

\caption[Spaghetti and density charts  for the variable serBilir]{Spaghetti and density charts  for the variable serBilir.}\label{fig:spaghetti}
\end{figure}
\end{Schunk}

The two charts thus created are shown in Figure \ref{fig:spaghetti}.

We can observe that some longitudinal covariates exhibit strong
skewness, as in the case of \texttt{serBilir}. Although in principle the
LMM can be used to model variables with skewed distributions, this may
sometimes lead to converge problems or poor model fit. It can thus be
advisable to transform such covariates to prevent these problems (but
note that this is not a required modelling choice, and one may
alternatively choose to avoid such transformation). For this reason, we
log-transform those longitudinal covariates with skewed distribution
before modelling them with LMMs:

\begin{Schunk}
\begin{Sinput}
ldata$logSerBilir = log(ldata$serBilir)
ldata$logSerChol = log(ldata$serChol)
ldata$logAlkaline = log(ldata$alkaline)
ldata$logSGOT = log(ldata$SGOT)
ldata$logProthrombin = log(ldata$prothrombin)
\end{Sinput}
\end{Schunk}

\hypertarget{model-estimation}{%
\subsection{Model estimation}\label{model-estimation}}

\hypertarget{step-1-estimation-of-the-mixed-effects-models}{%
\subsubsection{Step 1: estimation of the mixed-effects
models}\label{step-1-estimation-of-the-mixed-effects-models}}

The first step in the estimation of PRC involves modelling the evolution
over time of the longitudinal predictors through mixed-effects models.
Hereafter we model each of the longitudinal predictors using the LMM
given in equation \eqref{eq:lmmwithslope}, which comprises a random
intercept and a random slope for age. Estimation of this model for each
longitudinal predictor can be done using the function
\texttt{fit\_lmms}:

\begin{Schunk}
\begin{Sinput}
lmms = fit_lmms(y.names = c('logSerBilir', 'logSerChol', 'albumin', 'logAlkaline',
                            'logSGOT', 'platelets', 'logProthrombin'), 
                fixefs = ~ age, ranefs = ~ age | id, t.from.base = fuptime,
                long.data = ldata, surv.data = sdata, n.boots = 0, n.cores = 8, 
                verbose = FALSE)
\end{Sinput}
\end{Schunk}

The argument \texttt{y.names} is a character vector used to specify the
names of the longitudinal predictors in \texttt{ldata}. \texttt{fixefs}
and \texttt{ranefs} are formulas used to specify the fixed and random
effects part of the LMM using the \texttt{nlme} formula notation
\citep{pinheiro2022, galecki2013}. In the example,
\texttt{fixefs\ =\ \textasciitilde{}\ age} determines the inclusion of
the fixed effects part \(\beta_{s0} + \beta_{s1} a_{ij}\) of model
\eqref{eq:lmmwithslope}, and
\texttt{ranefs\ =\ \textasciitilde{}\ age\ \textbar{}\ id} the inclusion
of the random effects part \(u_{si0} + u_{si1} a_{ij}\) (allowing the
random intercept and random slopes to be correlated).

The arguments \texttt{long.data} and \texttt{surv.data} are used to
provide the names of the data frames containing the longitudinal
variables (\texttt{long.data}) and the survival data and baseline
covariates (\texttt{surv.data}) in the data formats described
previously. The argument \texttt{t.from.base} is used to specify the
name of the variable in \texttt{long.data} that contains the values of
time from baseline; this argument is used internally to check that the
data have been landmarked properly.

The \texttt{n.boots} argument is used to specify the number of bootstrap
samples to use for the CBOCP. For the time being we focus on model
estimation and on the prediction of the conditional probabilities,
setting \texttt{n.boots\ =\ 0}; later we will show how to compute the
CBOCP by setting \texttt{n.boots\ =\ 50}. Lastly, \texttt{n.cores}
allows to specify the number of cores to use to parallelize computations
(the default is \texttt{n.cores\ =\ 1}, i.e.~no parallelization), and
\texttt{verbose} is a logical value that indicates whether information
messages should be printed in the console (\texttt{TRUE}, default) or
not (\texttt{FALSE}). Additional arguments are described in the help
page, see \texttt{?fit\_lmms}.

The parameter estimates from the mixed models can be obtained from the
output of step 1 using \texttt{summary}. For example, to obtain the
parameters of the LMM for \texttt{albumin} we can use:

\begin{Schunk}
\begin{Sinput}
summary(lmms, yname = 'albumin', what = 'betas')
\end{Sinput}
\begin{Soutput}
#>  (Intercept)          age 
#>  3.822741450 -0.005710811
\end{Soutput}
\begin{Sinput}
summary(lmms, yname = 'albumin', what = 'variances')
\end{Sinput}
\begin{Soutput}
#> id = pdLogChol(age) 
#>             Variance     StdDev       Corr  
#> (Intercept) 8.864945e-02 0.2977405762 (Intr)
#> age         3.447615e-07 0.0005871639 -0.103
#> Residual    1.257161e-01 0.3545646671
\end{Soutput}
\end{Schunk}

From the output we can deduce that the ML estimates for the LMM
involving \texttt{albumin} are \(\hat{\beta} = (3.823, -0.0057)\),
\(\hat{\sigma}_{u0} = 0.2977\), \(\hat{\sigma}_{u1} = 0.00059\), and
\(\hat{\sigma}_{u0, u1} = -0.103 \cdot 0.2977 \cdot 0.00059\). The usual
table with parameter estimates, standard errors and p-values can be
obtained with

\begin{Schunk}
\begin{Sinput}
summary(lmms, yname = 'albumin', what = 'tTable')
\end{Sinput}
\begin{Soutput}
#>                    Value   Std.Error  DF   t-value       p-value
#> (Intercept)  3.822741450 0.106163343 566 36.008111 1.614057e-148
#> age         -0.005710811 0.002085903 566 -2.737812  6.379534e-03
\end{Soutput}
\end{Schunk}

\hypertarget{step-2-computation-of-the-predicted-random-effects}{%
\subsubsection{Step 2: computation of the predicted random
effects}\label{step-2-computation-of-the-predicted-random-effects}}

After the ML estimates from the LMM have been computed, the second step
of PRC involves the computation of the predicted random effects
\(\hat{u}_{si}\). Such computation can be performed using the function
\texttt{summarize\_lmms}:

\begin{Schunk}
\begin{Sinput}
pred_ranefs = summarize_lmms(object = lmms, n.cores = 8, verbose = FALSE)
summary(pred_ranefs)
\end{Sinput}
\begin{Soutput}
#> Number of predicted random effect variables: 14
#> Sample size: 278
\end{Soutput}
\end{Schunk}

Here, the \texttt{object} argument is used to pass the output of
\texttt{fit\_lmms} to \texttt{summarize\_lmms}; \texttt{n.cores} and
\texttt{verbose} are the same as in \texttt{fit\_lmms}. The output of
\texttt{summarize\_lmms} is a list that contains, among other elements,
a matrix with the predicted random effects called \texttt{ranef.orig},
where the row names display the subject identifiers:

\begin{Schunk}
\begin{Sinput}
round(pred_ranefs$ranef.orig[1:4, 1:4], 4)
\end{Sinput}
\begin{Soutput}
#>   logSerBilir_b_int logSerBilir_b_age logSerChol_b_int logSerChol_b_age
#> 2           -0.3830           -0.0017          -0.0712           0.0007
#> 3           -0.1171           -0.0006          -0.5985           0.0049
#> 4            0.1686            0.0009          -0.3704           0.0035
#> 5            0.3800            0.0012          -0.2910           0.0029
\end{Soutput}
\end{Schunk}

From the output we can deduce, for example, that the predicted random
intercept and random slope for \texttt{logSerBilir} and subject \(4\)
are \(\hat{u}_{1,0,4} = 0.1686\) and \(\hat{u}_{1,1,4} = 0.0009\).

\hypertarget{step-3-estimation-of-the-penalized-cox-model}{%
\subsubsection{Step 3: estimation of the penalized Cox
model}\label{step-3-estimation-of-the-penalized-cox-model}}

The last step in the estimation of PRC involves the estimation of model
\eqref{eq:cox} through PML. This can be achieved with the
\texttt{fit\_prclmm} function:

\begin{Schunk}
\begin{Sinput}
pencox = fit_prclmm(object = pred_ranefs, surv.data = sdata,
                    baseline.covs = ~ baselineAge + sex + treatment, penalty = 'ridge', 
                    standardize = TRUE, n.cores = 8, verbose = FALSE)
\end{Sinput}
\end{Schunk}

The \texttt{object} argument is used to pass the output of
\texttt{summarize\_lmms} to \texttt{fit\_prclmm}; \texttt{surv.data} is
the data frame that contains the information about survival data and
baseline covariates; \texttt{baseline.covs} is a formula used to define
which baseline covariates \(x_i\) should be included in model
\eqref{eq:cox} (with associated regression coefficient \(\gamma\)). The
\texttt{penalty} argument is a character that can take one of the
following values:

\begin{enumerate}
\def\labelenumi{\arabic{enumi}.}
\tightlist
\item
  \texttt{penalty\ =\ \textquotesingle{}ridge\textquotesingle{}} to
  estimate model \eqref{eq:cox} using the ridge or L2 penalty within the
  PML estimation;
\item
  \texttt{penalty\ =\ \textquotesingle{}lasso\textquotesingle{}} to
  estimate model \eqref{eq:cox} using the lasso or L1 penalty;
\item
  \texttt{penalty\ =\ \textquotesingle{}elnet\textquotesingle{}} to
  estimate model \eqref{eq:cox} using the elasticnet penalty
  \citep{zou2005}. If this penalty is chosen, additional arguments such
  as \texttt{n.alpha.elnet} and \texttt{n.folds.elnet} can be specified
  to determine how to select the additional tuning parameter
  (\(\alpha\)) used by this penalty through nested cross-validation.
\end{enumerate}

The \texttt{standardize} argument is used to determine whether the
predicted random effects should be standardized prior to inclusion in
the Cox model (default is \texttt{TRUE}). By default,
\texttt{fit\_prclmm} does not penalize baseline covariates, but this
default behaviour can be changed using the argument
\texttt{pfac.base.covs} argument (not shown here).

The \texttt{n.cores} and \texttt{verbose} arguments are the same as in
\texttt{fit\_lmms}. See \texttt{?fit\_prclmm} for a description of
further arguments.

The output of \texttt{fit\_prclmm} can be summarized through
\texttt{summary}:

\begin{Schunk}
\begin{Sinput}
summary(pencox)
\end{Sinput}
\begin{Soutput}
#> Fitted model: PRC-LMM
#> Penalty function used: ridge
#> Tuning parameters:
#>      lambda alpha
#> 1 0.2126761     0
#> Sample size: 278
#> Number of events: 107
#> Bootstrap optimism correction: not computed
#> Penalized likelihood estimates (rounded to 4 digits):
#>   baselineAge sexfemale treatmentD-penicil logSerBilir_b_int logSerBilir_b_age
#> 1      0.0476   -0.2872            -0.0157            0.4341          111.3935
#>   logSerChol_b_int logSerChol_b_age albumin_b_int albumin_b_age
#> 1           0.0986         -10.5311       -1.1361      23070.92
#>   logAlkaline_b_int logAlkaline_b_age logSGOT_b_int logSGOT_b_age
#> 1            0.0874          -12.5617         0.238       272.246
#>   platelets_b_int platelets_b_age logProthrombin_b_int logProthrombin_b_age
#> 1         -0.0011         -0.2046               2.8114            -573.3093
\end{Soutput}
\end{Schunk}

The PML estimates of \(\gamma\) and \(\delta\) can be obtained through

\begin{Schunk}
\begin{Sinput}
step3summary = summary(pencox)
ls(step3summary)
\end{Sinput}
\begin{Soutput}
#> [1] "coefficients" "data_info"    "model_info"   "tuning"
\end{Soutput}
\begin{Sinput}
step3summary$coefficients
\end{Sinput}
\begin{Soutput}
#>   baselineAge sexfemale treatmentD-penicil logSerBilir_b_int logSerBilir_b_age
#> 1   0.0475985 -0.287243        -0.01567369         0.4340672          111.3935
#>   logSerChol_b_int logSerChol_b_age albumin_b_int albumin_b_age
#> 1        0.0986092        -10.53106      -1.13607      23070.92
#>   logAlkaline_b_int logAlkaline_b_age logSGOT_b_int logSGOT_b_age
#> 1        0.08741668         -12.56169     0.2379553       272.246
#>   platelets_b_int platelets_b_age logProthrombin_b_int logProthrombin_b_age
#> 1    -0.001122804      -0.2046088             2.811446            -573.3093
\end{Soutput}
\end{Schunk}

\hypertarget{computing-predictions}{%
\subsection{Computing predictions}\label{computing-predictions}}

The function \texttt{survpred\_prclmm} can be used to compute the
conditional survival probabilities
\(\hat{S}_i(t | t_L, x_i, \mathcal{Y}_i(t_L)),\) \(t \geq t_L\) in
\eqref{eq:predsurv}. For the subjects that have been used to estimate
PRC, such computation can be performed using \texttt{survpred\_prclmm}
as follows:

\begin{Schunk}
\begin{Sinput}
preds = survpred_prclmm(step1 = lmms, step2 = pred_ranefs, step3 = pencox, times = 3:7)
\end{Sinput}
\end{Schunk}

The \texttt{step1}, \texttt{step2} and \texttt{step3} arguments are used
to pass the outputs of the 3 estimation steps to the function; the
\texttt{times} argument is a vector with the prediction times at which
one wishes to evaluate the conditional survival probabilities. The
predicted survival probabilities are stored in the
\texttt{predicted\_survival} element of the function output:

\begin{Schunk}
\begin{Sinput}
ls(preds)
\end{Sinput}
\begin{Soutput}
#> [1] "call"               "predicted_survival"
\end{Soutput}
\begin{Sinput}
head(preds$predicted_survival)
\end{Sinput}
\begin{Soutput}
#>   id      S(3)      S(4)      S(5)      S(6)      S(7)
#> 2  2 0.9398512 0.8867592 0.8329966 0.7813647 0.7008304
#> 3  3 0.8555809 0.7392053 0.6316391 0.5377700 0.4090876
#> 4  4 0.8138498 0.6709525 0.5451302 0.4407852 0.3071611
#> 5  5 0.9460431 0.8981124 0.8492655 0.8020394 0.7277052
#> 6  6 0.9383339 0.8839878 0.8290416 0.7763599 0.6943714
#> 7  7 0.9718724 0.9462254 0.9193945 0.8927316 0.8491681
\end{Soutput}
\end{Schunk}

The function \texttt{survplot\_prc} allows to visualize predictions for
a sample of individuals:

\begin{Schunk}
\begin{Sinput}
survplot_prc(step1 = lmms, step2 = pred_ranefs, step3 = pencox, 
             ids = c(54, 111, 173, 271), tmax = 12)
\end{Sinput}
\begin{figure}
\includegraphics{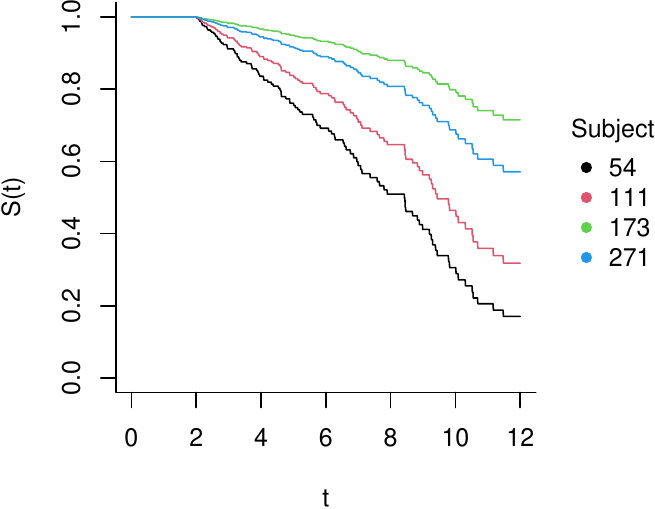} \caption{Predicted survival probabilities $\hat{S}_i(t | 2), \:t \in [2, 12]$ for subjects $i \in \{54, 111, 173, 271\}$.}\label{fig:survplot}
\end{figure}
\end{Schunk}

The \texttt{ids} argument is used to indicate the subjects for whom the
curve should be displayed, and \texttt{tmax} sets the upper limit for
the \(x\) axis. The chart, displayed in Figure \ref{fig:survplot}, shows
the predicted survival probability \(\hat{S}_i(t | 2)\) for subjects 54,
111, 173, and 271. Notice that the survival probability up to the 2 year
landmark is 1 because our modelling approach conditions on being still
at risk at the landmark time.

Prediction for new subjects that have not been used for model estimation
is a bit more involved, as it additionally requires to compute the
predicted random effects for the new subjects using equation
\eqref{eq:predranef} based on the parameter estimates obtained in the
first step of PRC. For illustration purposes, suppose that we have data
from 3 subjects stored in two data frames called \texttt{new\_ldata} and
\texttt{new\_sdata}:

\begin{Schunk}
\begin{Sinput}
new_ldata = subset(ldata, id %in% c(5, 54, 110))
new_sdata = subset(sdata, id %in% c(5, 54, 110), c('id', 'baselineAge', 'sex', 'treatment'))
head(new_ldata)
\end{Sinput}
\begin{Soutput}
#>     id      age   fuptime serBilir serChol albumin alkaline  SGOT platelets
#> 23   5 38.10645 0.0000000      3.4     279    3.53      671 113.2       136
#> 24   5 38.65130 0.5448472      1.9      NA    3.28      689 103.9       114
#> 25   5 39.17698 1.0705290      2.5      NA    3.34      652 117.8        99
#> 419 54 39.19888 0.0000000      1.3     288    3.40     5487  73.5       254
#> 420 54 39.69171 0.4928266      1.5      NA    3.22     1580  71.3       112
#> 421 54 40.19001 0.9911291      2.6      NA    3.48     2127  86.8       207
#>     prothrombin logSerBilir logSerChol logAlkaline  logSGOT logProthrombin
#> 23         10.9   1.2237754   5.631212    6.508769 4.729156       2.388763
#> 24         10.7   0.6418539         NA    6.535241 4.643429       2.370244
#> 25         10.5   0.9162907         NA    6.480045 4.768988       2.351375
#> 419        11.0   0.2623643   5.662960    8.610137 4.297285       2.397895
#> 420        18.0   0.4054651         NA    7.365180 4.266896       2.890372
#> 421        10.9   0.9555114         NA    7.662468 4.463607       2.388763
\end{Soutput}
\begin{Sinput}
head(new_sdata)
\end{Sinput}
\begin{Soutput}
#>      id baselineAge    sex treatment
#> 23    5    38.10645 female   placebo
#> 419  54    39.19888 female D-penicil
#> 859 110    38.91140 female D-penicil
\end{Soutput}
\end{Schunk}

Note that the variables and their variable type in \texttt{new\_ldata}
and \texttt{new\_sdata} should be the same as in the \texttt{ldata} and
\texttt{sdata}; the only exception to this is that \texttt{new\_sdata}
does not need to contain information about survival, so unlike
\texttt{sdata} it does not comprise the \texttt{time} and \texttt{event}
variables.

To compute predicted probabilities for new subjects, it is once again
possible to resort to \texttt{survpred\_prclmm}; now, it is necessary to
specify the arguments \texttt{new.longdata} and \texttt{new.basecovs} to
supply the data about the new subjects to \texttt{survpred\_prclmm}:

\begin{Schunk}
\begin{Sinput}
pred_new = survpred_prclmm(step1 = lmms, step2 = pred_ranefs, step3 = pencox, times = 3:7,
                           new.longdata = new_ldata, new.basecovs = new_sdata)
pred_new$predicted_survival
\end{Sinput}
\begin{Soutput}
#>      id      S(3)      S(4)      S(5)      S(6)      S(7)
#> 5     5 0.9460431 0.8981124 0.8492655 0.8020394 0.7277052
#> 54   54 0.9115188 0.8357020 0.7611779 0.6918065 0.5880763
#> 110 110 0.9505706 0.9064581 0.8612933 0.8174139 0.7478898
\end{Soutput}
\end{Schunk}

\hypertarget{evaluation-of-the-predictive-performance-1}{%
\subsection{Evaluation of the predictive
performance}\label{evaluation-of-the-predictive-performance-1}}

As explained in the Statistical Methods Section, estimation of the
predictive performance in \CRANpkg{pencal} is done through a CBOCP that
allows to obtain unbiased estimates of predictive performance as
measured by the tdAUC, C index and Brier score.

Before computing the (potentially time-consuming) CBOCP, one may want to
first have a look at the naïve (biased) estimates of predictive
performance. This can be done using the function
\texttt{performance\_prc}:

\begin{Schunk}
\begin{Sinput}
# naive performance (biased, optimistic estimate)
naive_perf = performance_prc(step2 = pred_ranefs, step3 = pencox, metric = 'tdauc',
                             times = 3:7, n.cores = 8, verbose = FALSE)
\end{Sinput}
\begin{Soutput}
#> Warning in performance_prc(step2 = pred_ranefs, step3 = pencox, metric =
#> "tdauc", : The cluster bootstrap optimism correction has not been performed
#> (n.boots = 0). Therefore, only the apparent values of the performance values
#> will be returned.
\end{Soutput}
\end{Schunk}

Here \texttt{pred\_ranefs} is the output of step2 of PRC and
\texttt{pencox} the output of step 3. The \texttt{metric} argument can
be used to specify the performance measures to be computed (possible
values are \texttt{tdauc}, \texttt{c} and \texttt{brier}), whereas the
\texttt{times} argument is used to specify the time points at which the
tdAUC and Brier score should be evaluated (\(t = 3, 4, 5, 6, 7\) in this
example). Notice that when \texttt{fit\_lmms} has been run with
\texttt{n.boots\ =\ 0}, \texttt{performance\_prc} returns a warning to
inform users that the CBOCP has not been performed; for now, we can
ignore this warning (which we will address soon by refitting PRC with
\texttt{n.boots\ =\ 50}).

\begin{Schunk}
\begin{Sinput}
naive_perf
\end{Sinput}
\begin{Soutput}
#> $call
#> performance_prc(step2 = pred_ranefs, step3 = pencox, metric = "tdauc", 
#>     times = 3:7, n.cores = 8, verbose = FALSE)
#> 
#> $tdAUC
#>   pred.time tdAUC.naive optimism.correction tdAUC.adjusted
#> 1         3      0.9439                  NA             NA
#> 2         4      0.9351                  NA             NA
#> 3         5      0.9266                  NA             NA
#> 4         6      0.8981                  NA             NA
#> 5         7      0.8831                  NA             NA
\end{Soutput}
\end{Schunk}

From the output we can observe that the naïve estimate of the tdAUC
ranges from 0.9439 for predictions of survival at \(t = 3\) up to 0.8831
for predictions at \(t = 7\). These naïve (in-sample) measurements of
predictive performance may be optimistically biased due to overfitting,
i.e., the fact that they are evaluated using the same data on which PRC
was estimated. To correct for this potential source of bias, below we
show how to implement the CBOCP to obtain unbiased estimates of the
tdAUC and C index.

Computation of the CBOCP requires to repeat the 3 estimation steps of
PRC for each bootstrap samples; this can be done by rerunning the
functions \texttt{fit\_lmms}, \texttt{summarize\_lmms} and
\texttt{fit\_prclmm} with the same arguments used previously, but
setting \texttt{n.boots} within \texttt{fit\_lmms} to an integer value
larger than 0. \texttt{n.boots} specifies the number of bootstrap
samples to use to compute the CBOCP. In the example below we set
\texttt{n.boots\ =\ 50} (note that larger values of \texttt{n.boots} can
increase the accuracy of the CBOCP estimates, but at the same time they
increase computing time).

\begin{Schunk}
\begin{Sinput}
step1 = fit_lmms(y.names = c('logSerBilir', 'logSerChol', 'albumin', 'logAlkaline',
                             'logSGOT', 'platelets', 'logProthrombin'), 
                fixefs = ~ age, ranefs = ~ age | id, t.from.base = fuptime,
                long.data = ldata, surv.data = sdata, n.boots = 50, n.cores = 8, 
                verbose = FALSE)
step2 = summarize_lmms(object = step1, n.cores = 8, verbose = FALSE)
step3 = fit_prclmm(object = step2, surv.data = sdata,
                   baseline.covs = ~ baselineAge + sex + treatment,
                   penalty = 'ridge', n.cores = 8, verbose = FALSE)
\end{Sinput}
\end{Schunk}

Once all computations are finished, it suffices to supply the refitted
outputs of step 2 and step 3 to \texttt{performance\_prc}:

\begin{Schunk}
\begin{Sinput}
# bootstrap-corrected performance (unbiased estimate)
cbocp = performance_prc(step2 = step2, step3 =  step3, metric = c('tdauc', 'brier'), 
                        times = 3:7, n.cores = 8, verbose = FALSE)
cbocp
\end{Sinput}
\begin{Soutput}
#> $call
#> performance_prc(step2 = step2, step3 = step3, metric = c("tdauc", 
#>     "brier"), times = 3:7, n.cores = 8, verbose = FALSE)
#> 
#> $tdAUC
#>   pred.time tdAUC.naive optimism.correction tdAUC.adjusted
#> 1         3      0.9439             -0.0063         0.9376
#> 2         4      0.9351             -0.0153         0.9198
#> 3         5      0.9266             -0.0133         0.9133
#> 4         6      0.8981             -0.0091         0.8890
#> 5         7      0.8831             -0.0121         0.8710
#> 
#> $Brier
#>   pred.time Brier.naive optimism.correction Brier.adjusted
#> 1         3      0.0571              0.0151         0.0722
#> 2         4      0.0699              0.0281         0.0980
#> 3         5      0.0844              0.0340         0.1184
#> 4         6      0.0953              0.0364         0.1317
#> 5         7      0.1007              0.0434         0.1441
\end{Soutput}
\end{Schunk}

In the outputs above, the columns \texttt{tdAUC.naive} and
\texttt{Brier.naive} contain the naïve estimates of the tdAUC and Brier
score; \texttt{optimism.correction} reports the values of the estimated
optimism correction from the CBOCP for the two metrics; finally,
\texttt{tdAUC.adjusted} and \texttt{Brier.adjusted} contain the unbiased
estimates of the tdAUC and Brier score.

As expected, the unbiased estimates of predictive performance are
somewhat worse than the naïve ones. For example, the tdAUC estimate for
predictions at \(t = 3\) is 0.9376 instead of the naïve estimate 0.9439.
Similarly, the Brier score estimate for predictions at \(t = 3\) is
0.0722 instead of the naïve estimate 0.0571.

\hypertarget{evaluation-of-computing-time}{%
\section{Evaluation of computing
time}\label{evaluation-of-computing-time}}

We now turn our attention to the relationship between the sample size
\(n\), number of longitudinal covariates \(p\) and number of bootstrap
replicates \(B\) on computing time. Furthermore, we look into how
parallel computing may be used to reduce computing time for the CBOCP.
To gain insight into these relationships, we simulate data from the PRC
LMM model using the function \texttt{simulate\_prclmm\_data} according
to four simulation scenarios:

\begin{itemize}
\tightlist
\item
  in simulation 1 we study the effect of \(n\) on the estimation of PRC.
  To this aim, we let \(n \in \{100, 200, 400, 600, 800, 1000\}\) and
  fix \(p = 10\);
\item
  in simulation 2 we study the effect of \(p\) on the estimation of PRC
  by taking \(p \in \{5, 10, 20, 30, 40, 50\}\) and fixing \(n = 200\);
\item
  in simulation 3 we shift our attention to the effect of \(B\) on the
  computing time of the CBOCP. We let
  \(B \in \{50, 100, 200, 300, 400, 500\}\), fixing \(n = 200\) and
  \(p = 10\);
\item
  finally, in simulation 4 we compute PRC and the CBOCP on a dataset
  where \(n = 200\), \(p = 50\) and \(B = 50\) using an increasing
  number of cores, namely \(\{1, 2, 3, 4, 8, 16\}\).
\end{itemize}

Computations were performed on an AMD EPYC 7662 processor with 2 GHz
CPU, using a single core for simulations 1, 2 and 3, and a number of
cores ranging from 1 to 16 in simulation 4. Computing time was measured
using the \texttt{rbenchmark} package \citep{kusnierczyk2012}.

\begin{Schunk}
\begin{figure}[htbp]

{\centering \includegraphics[width=5.5in]{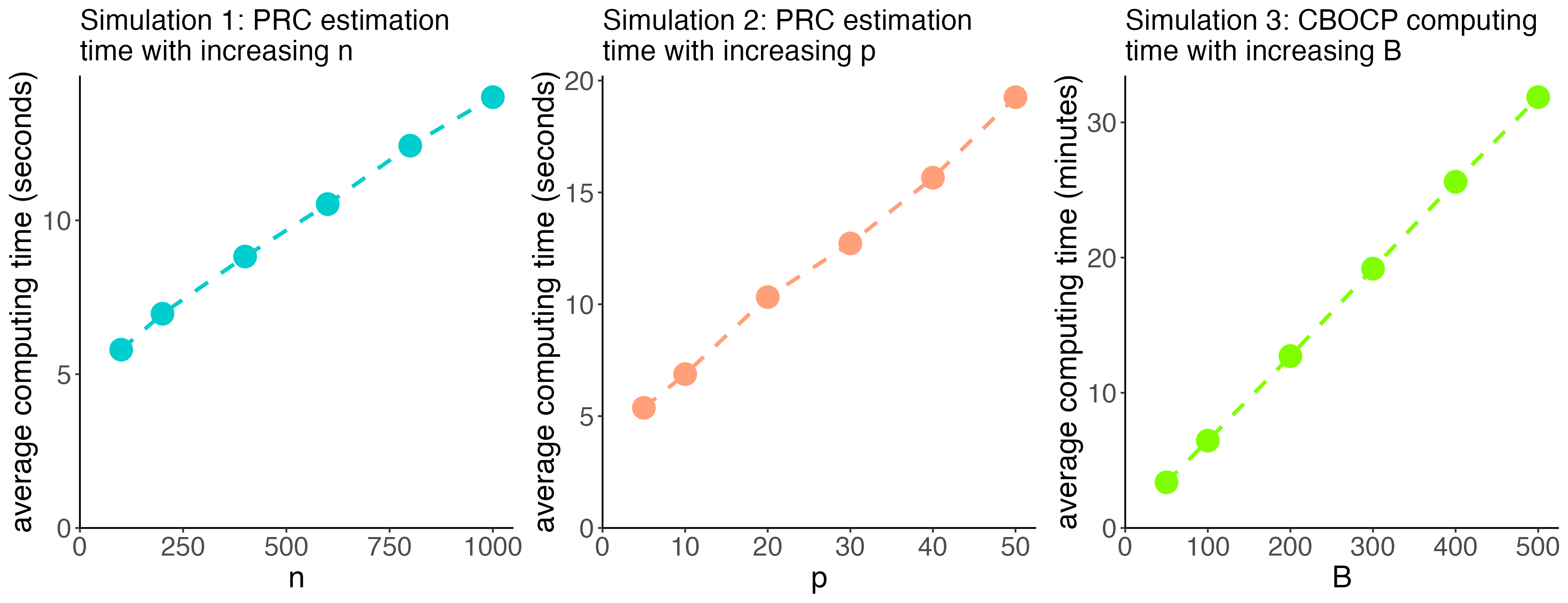} 

}

\caption[Left and center]{Left and center: average computing time (in seconds) of the estimation of the PRC LMM model as a function of the sample size $n$ (simulation 1, left) and of the number of longitudinal predictors $p$ (simulation 2, center). Right: average computing time (in minutes) for the computation of the CBOCP as a function of $B$ (simulation 3).}\label{fig:ctime1}
\end{figure}
\end{Schunk}

The charts in Figure \ref{fig:ctime1} display the average computing time
over 10 replications of simulations 1, 2 and 3. In simulation 1, the
average computing time increases from 5.80 seconds when \(n = 100\) to
14.01 seconds when \(n = 1000\), so we observe a 2.4-fold increase in
computing time as \(n\) increases by a factor of 10. In simulation 2,
the average computing time increases from 5.37 seconds when \(p = 5\) to
19.26 seconds when \(p = 50\), yielding a 3.6-fold increase in computing
time as \(p\) increases by a factor of 10. Lastly, in simulation 3 the
average computing time increases from 3.38 minutes when \(B = 50\) to
32.32 minutes when \(B = 500\), with a 9.6-fold time increase
corresponding to a 10-fold increase in \(B\).

Overall, the results of simulations 1 and 2 indicate that computing time
for the estimation of PRC increases linearly with, but less than
proportionally to, \(n\) and \(p\) (meaning that a \(k\)-fold increase
in \(n\) or \(p\) will typically increase computing time by a factor
smaller than \(k\)). Instead, from simulation 3 we can observe that the
computing time of the CBOCP increases proportionally to \(B\).

The results of simulations 1, 2 and 3 show that whereas the estimation
of the PRC model itself typically requires only a few seconds, the
computation of the CBOCP is more intensive and can require several
minutes. This is due to the fact that the CBOCP requires the PRC
modelling steps to be repeated on each bootstrap sample, effectively
requiring to compute PRC \(B + 1\) times (once on the original dataset +
\(B\) times on the \(B\) bootstrap datasets). To reduce the computing
time needed to compute the CBOCP, \texttt{pencal} enables users to
easily parallelize computations through the argument \texttt{n.cores}
within \texttt{fit\_lmms}, \texttt{summarize\_lmms} and
\texttt{fit\_prclmm}. In simulation 4, we show the effect that
increasing the number of cores has on the computing time of the CBOCP.

\begin{Schunk}
\begin{figure}[htbp]

{\centering \includegraphics[width=4.5in]{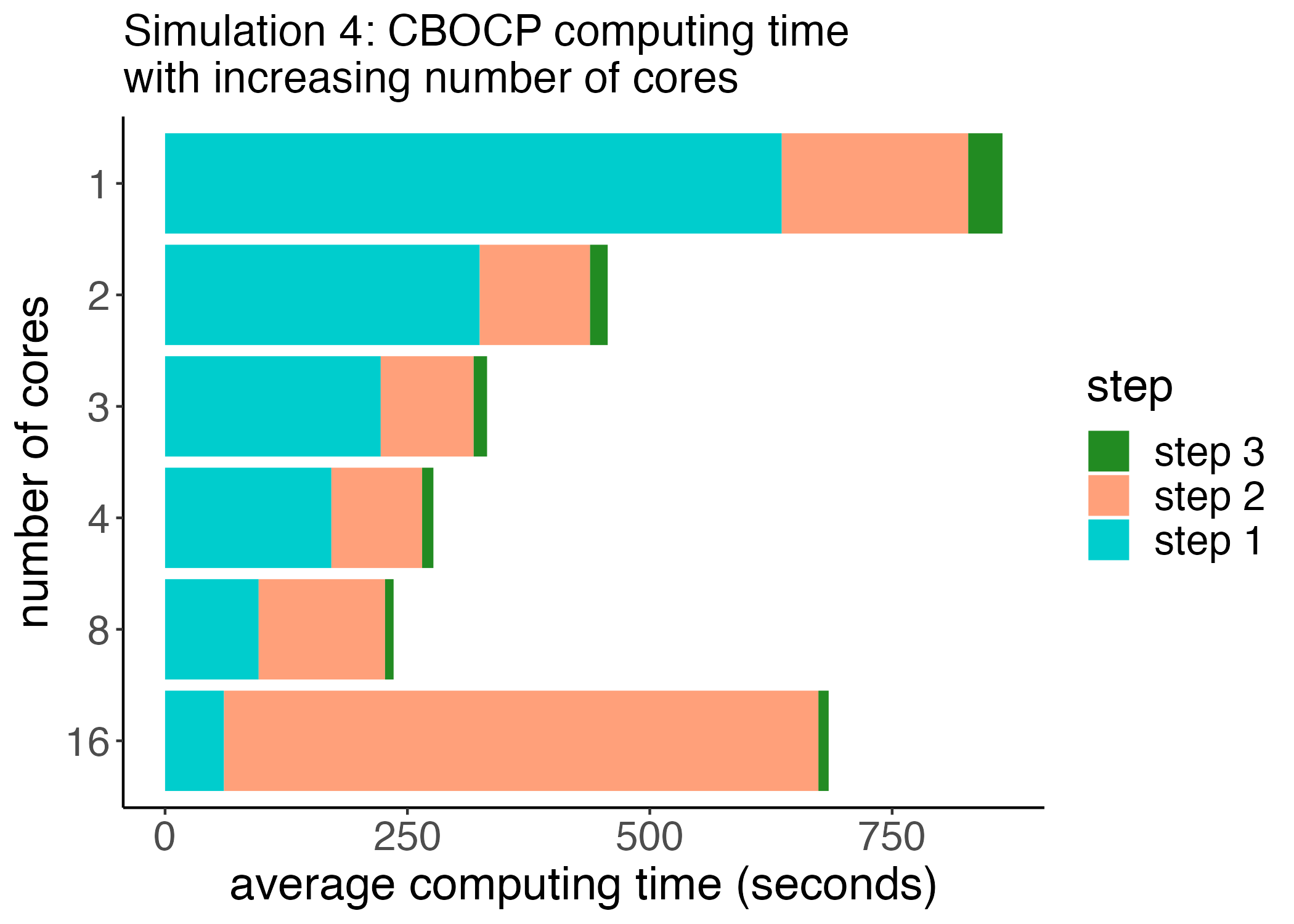} 

}

\caption[Average computing time (in seconds) for the estimation of the PRC LMM model and the computation of the CBOCP as a function of the number of cores]{Average computing time (in seconds) for the estimation of the PRC LMM model and the computation of the CBOCP as a function of the number of cores.}\label{fig:ctime4}
\end{figure}
\end{Schunk}

Figure \ref{fig:ctime4} shows the average computing time of the CBOCP
over 10 replications of simulation 4. When looking at the total
computing time, we can see that increasing the number of cores from 1 to
8 progressively decreases computing time, reducing it from 863.8 seconds
without parallelization up to 235.7 seconds when using 8 cores (-68\%).
The most significant time gains are from 1 to 2 cores (-407.2 seconds)
and then from 2 to 3 (-124.4 seconds). Interestingly, further doubling
the number of cores from 8 to 16 proves to be detrimental, increasing
computing time from 235.7 to 684.5 seconds.

To understand this initially decreasing, but later increasing pattern,
it is useful to consider the computing time of each of the modelling
steps separately. By looking at Figure \ref{fig:ctime4} we notice that
step 1, which involves the estimation of \(p \cdot (B+1)\) LMMs, is the
most time-consuming step; its computing time consistently decreases as
the number of cores decreases (from 636.1 to 60.8 seconds).

The same does not apply to step 2, where computing time decreases from 1
to 4 cores (from 192.3 to 93.6 seconds), but then increases considerably
when going from 8 (130.4 seconds) to 16 cores (613.2 seconds). This
pattern is primarily due to the fact that step 2 mostly involves simple
linear algebra: parallelizing this step on a large number of cores may
be detrimental, as the (limited) time gain that can be achieved by doing
these simple computations in parallel may be more than compensated by
the time cost of dispatching the necessary matrices and vectors to many
cores and recombining the results at the end of the parallelization.

As concerns step 3, we can see that it is the lightest step in terms of
computing time. The pattern is consistently decreasing from 1 (35.4
seconds) to 8 cores (8.8 seconds), with a slight increase when using 16
cores (10.6 seconds).

In conclusion, the results of simulation 4 show how it may be advisable
to parallelize computations to compute the CBOCP, but without using an
excessive number of cores (specially for step 2). Our general advice is
to use between 3 and 8 cores for optimal performance, nevertheless we
emphasize how the effect of the number of cores on computing time may
differ from the patterns in Figure \ref{fig:ctime4} depending on a
combination of factors such as \(n\), \(p\), number of repeated
measurements per subject, and \(B\). Furthermore, we notice that within
\texttt{pencal}, a different number of cores can be chosen for each
modelling step; in the example of simulation 4, the optimal performance
would be achieved using 16 cores for step 1, 4 cores for step 2, and 8
cores for step 3 (but using 8 cores for all 3 steps isn't much less
efficient).

\hypertarget{summary-and-discussion}{%
\section{Summary and discussion}\label{summary-and-discussion}}

The \texttt{R} package \CRANpkg{pencal} provides a user-friendly
implementation of Penalized Regression Calibration (PRC,
\citet{signorelli2021}), a statistical method that can be used to
implement dynamic prediction of time-to-event outcomes in longitudinal
studies where both time-independent and longitudinal (i.e.,
time-dependent) covariates are available as possible predictors of
survival. The package comprises functions for the estimation of PRC and
the prediction of survival, as well as functions to compute unbiased
estimates of predictive performance through a cluster bootstrap
procedure. Because computing such bootstrap procedure may be
time-consuming, the package automatically parallelizes repetitive
computations using the \texttt{\%dopar\%} operator from the
\CRANpkg{foreach} package \citep{foreach2022}.

\CRANpkg{pencal} focuses on problems where a single survival outcome is
measured with right-censoring. As such, it is not designed to handle
interval censoring or competing risks. The modelling of the longitudinal
covariates is performed using either the LMM or the MLPMM, which are
linear models that are mostly suitable for the analysis of continuous
outcomes. Implementing generalized linear mixed models (GLMMs) would
make it possible to properly deal with binary and discrete longitudinal
covariates, however the estimation of GLMMs and the computation of the
predicted random effects are more time-consuming and more prone to
convergence problems, two aspects that would particularly complicate the
computation of the CBOCP. For this reason we did not pursue GLMMs
further, but leave them as a topic of future research. Users dealing
with discrete longitudinal covariates may consider log-transforming them
before modelling with a LMM within \CRANpkg{pencal} (specially if such
covariates are right-skewed and/or exhibit overdispersion). Despite this
latter limitation, a recent benchmarking study showed that PRC
outperformed several alternative modelling approaches when applied to
multiple real-world datasets \citep{signorelli2024}.

Two modelling choices deserve particular attention when implementing PRC
in specific application contexts. The first refers to the choice of the
covariates to include in the fixed and random effects parts of the LMM
of Equation \eqref{eq:lmm}. In principle, one may want to model the
response variable as flexibly as possible, including several fixed
effect covariates and multiple random effects in the LMM. However, when
doing this one should consider that the purpose of the LMM is to provide
subject-specific summaries of the individual trajectories. Thus,
\emph{the primary goal of the LMM in step 1 is that of obtaining
predicted random effects that are good summaries of how the trajectory
of a given subject differs from the population average}. In practice,
such purpose may be more easily achieved using a simple mixed model that
allows for a clear interpretation of its random effects, rather than
using a complex one where the interpretation of each random effect may
be unclear / complicated. The LMM of equation \eqref{eq:lmmwithslope}
(or, alternatively, the same model with follow-up time included as
covariate instead of age) is a good example of simple LMM with clearly
interpretable random effects, as the random intercept allows to
distinguish subjects with high and low initial levels of the covariate,
and the random slope to identify subjects with faster and slower
progression rates. Therefore, even though \texttt{fit\_lmms} makes it
possible to consider complex fixed and random effects formulas, we still
advise users to consider simpler mixed models in step 1 (and to compare
the predictive performance of PRC using either approach, eventually
choosing the approach that delivers more accurate predictions if there
is a substantial difference).

A second important modelling choice when using \CRANpkg{pencal} is which
penalty function should be used in step 3. In general, this is a
modelling choice that may dependent on the specific application and its
features (sample size, number of predictors and number of available
repeated measurements per subject). \citet{signorelli2021} performed
several simulation studies focused on situations with small sample sizes
(\(n = 100\) and \(n = 300\)) and sparse data generating processes for
the survival outcome, whose results showed that the ridge and elasticnet
penalty yielded better performance than the lasso penalty. Our
experience is that in general the ridge penalty may be preferable both
to elasticnet and the lasso in scenarios with small or moderate sample
sizes, where little information is available to estimate the \(\alpha\)
tuning parameter of elasticnet or to reliably perform variable selection
with the lasso. Beyond this, it is always possible to use a data-driven
approach to choose which penalty to use by estimating PRC using the 3
different penalties and comparing how this affects predictive
performance.

\hypertarget{software-and-code-availability}{%
\subsection{Software and code
availability}\label{software-and-code-availability}}

The \texttt{R} package \CRANpkg{pencal} can be downloaded from CRAN at
\href{https://cran.r-project.org/package=pencal}{cran.r-project.org/package=pencal}.
The development version of the package is available on Github at
\href{https://github.com/mirkosignorelli/pencal_devel}{github.com/mirkosignorelli/pencal\_devel}.
The code used in the simulations for the evaluation of computing time is
available at\\
\href{https://github.com/mirkosignorelli/pencal_sims/}{github.com/mirkosignorelli/pencal\_sims/}.

\hypertarget{acknowledgements}{%
\subsection{Acknowledgements}\label{acknowledgements}}

The author kindly acknowledges funding from the Netherlands eScience
Center Fellowship Programme.

\bibliography{pencal-RJ.bib}

\address{%
Mirko Signorelli\\
Mathematical Institute, Leiden University\\%
Niels Bohrweg 1, 2333 CA Leiden (NL)\\
\url{https://mirkosignorelli.github.io}\\%
\textit{ORCiD: \href{https://orcid.org/0000-0002-8102-3356}{0000-0002-8102-3356}}\\%
\href{mailto:msignorelli.papers@gmail.com}{\nolinkurl{msignorelli.papers@gmail.com}}%
}

\end{article}

\end{document}